\documentclass[aps,prl,showpacs,twocolumn]{revtex4-1}
\usepackage{graphicx}
\usepackage[usenames]{color}
\usepackage{natbib}

\newcommand{\fig}[1]{Fig.~\ref{#1}}
\newcommand{\be}[1]{\begin{equation}\label{#1}}
\newcommand{\ee}{\end{equation}}
\begin{document}

\title{Two-electron effects  pervading the formation of highly excited neutral atoms in the break-up of strongly driven H$_{2}$}

\author{A. Emmanouilidou$^{1,2}$\email{}, D. Tchitchekova$^2$ C. Lazarou$^{1}$, and U. Eichmann$^{3,4}$}

\affiliation{
$^1$ Department of Physics and Astronomy, University College London, Gower Street, London WC1E 6BT, United Kingdom\\
$^2$Chemistry Department, University of Massachusetts at Amherst, Amherst, Massachusetts, 01003, U.S.A.\\
$^3$ Max-Born-Institute, Max-Born-Strasse 2a, 12489 Berlin, Germany. \\
$^4$Institut f\"ur Optik und Atomare Physik, Technische Universit\"at Berlin, 10632 Berlin, Germany.
}

\begin{abstract}
We present the first theoretical treatment of the formation of
highly excited neutral H atoms (H$^{*}$) for strongly driven
H$_{2}$. This process, with one electron ionizing and one captured
in a Rydberg state, was recently reported in an experimental study
in Phys. Rev. Lett {\bf 102}, 113002 (2009). 
We show that  two
pathways underlie this process: one mostly of non-sequential nature---more so for small intensities---resembling
non-sequential double ionization, and one mostly of sequential nature---more so for high intensities---resembling
double ionization through enhanced ionization. We also predict a
new feature, asymmetric energy sharing between H$^{*}$ and H$^{+}$
with increasing intensity. This feature is a striking demonstration
of the influence the electron has on the nuclear motion.

  \end{abstract}
\date{\today}

\maketitle

A wealth of physical phenomena is manifested during fragmentation of
molecules driven by intense infrared laser fields. In the case of
H$_{2}$---the most fundamental diatomic molecule---such phenomena
include, for instance, non-sequential double ionization (NSDI)
\cite{Alnaser,Niikura} with the re-collision (three-step) model
\cite{Corkum} underlying this process; double ionization through
enhanced ionization (EIDI) with one electron escaping early on while
the other ionizes later at intermediate distances of the nuclei of
the remaining molecular ion \cite{bandrauk1996,
Seideman,Villeneuve}. Very recently, formation of highly excited
neutral H atoms  \cite{Eichmann} was identified as yet another
interesting phenomenon in driven H$_{2}$. It was attributed to the
``frustrated tunnel ionization" mechanism leading to the recapture
of an electron. First seen in atoms \cite{Eichmann2} this mechanism
 appears to be
 a general one underlying the
formation of excited fragments during Coulomb explosion in strongly
driven molecules
\cite{nubbe}.
However, a full theoretical treatment and understanding is pending.
This is partially due to the fact that theoretical studies
including both nuclear and electronic motion in strongly driven
systems are computationally very demanding. Most studies fix the
nuclei and focus on electronic motion \cite{Saenz2006, Alnaser} 
or ignore the electronic continuum and study nuclear motion
\cite{nuclear}; in a recent refined study the electronic structure
is used as input to study the energy spectra of the Coulomb
exploding nuclei \cite{Madsen}.

%in the molecular ion the one electron ionization rate is strongly enhanced at an intermediate distance of the
% nuclei due to the electron localization on the upper well and subsequent tunneling and escape through the barrier that separates the two wells.

  In this Letter, we present the first
comprehensive theoretical study on the  break-up of strongly driven
H$_{2}$. We use a three-dimensional semi-classical approach that
explicitly accounts for the Coulomb singularity.
We focus on the Coulomb explosion
process involving a highly excited neutral hydrogen atom (H$^{*}$)
and a proton. The difficulty in its theoretical treatment lies in
the fact that, as we show,
two-electron effects pervade the two pathways underlying the formation
of H$^{*}$. This is in contrast to one-electron effects pervading 
the well known enhanced ionization process. Moreover, correctly accounting for this process requires treating
the electronic and nuclear motion on an equal footing. By doing so, we identify a transition with increasing laser intensity
from symmetric to asymmetric energy sharing between H$^{+}$ and
H$^{*}$.  We show this asymmetry in energy sharing to be a striking
signature of how the motion of a light
particle---electron---influences the motion of heavy ones---nuclei. This questions whether this asymmetry could be predicted by the widely used in quantum mechanics Born-Oppenheimer (BO) approximation, since the latter
 assumes that the massive nuclei are nearly fixed with respect to electron motion.

{\it Model}. Our calculations are performed in a classical framework
with the Coulomb singularity explicitly accounted for. Previously,
we successfully used a three-dimensional classical model for fixed
nuclei diatomics to predict new phenomena: an antiparallel electron
escape for intermediate laser field intensities in the
over-the-barrier regime \cite{Emmanouilidou2009}. The current model extends our previous technique
to also account for nuclear motion.

Our model entails setting up the initial phase space distribution.
The electronic initial state is detailed in
\cite{Emmanouilidou2009} both for the tunneling and the
over-the-barrier regime. We use 0.57 a.u. and 1.28 a.u. as the first
and second ionization potentials, respectively. We take the initial
vibrational state of the nuclei to be the ground state,
$E_{0}\approx0.01$ a.u, of the Morse potential $V_{M}(R)=D
(1-e^{-\beta (R-R_{0})})^2$ with $D=0.174$ a.u., $\beta=1.029$ a.u.
\cite{WignerMorse} and R the inter-nuclear distance. We choose the
Wigner distribution to describe the initial state of the nuclei due
to the good agreement with the position and  momentum distributions
of the initial quantum mechanical state.  (For the Wigner function
of the ground state Morse potential see \cite{WignerMorse}). We
consider an intensity in the tunneling regime of 1.5$\times 10^{14}$
Watts/cm$^2$ and an intermediate intensity in the over-the-barrier
regime of 2.5$\times 10^{14}$ Watts/cm$^2$. For this latter
intensity, for the majority of trajectories the electronic state is
initiated with the tunneling model, thus the term intermediate. Both
intensities are high enough to justify restricting the initial
distance of the nuclei to the equilibrium distance of the ground
vibrational state, $R_{0}=1.4$ a.u. \cite{Saenz}.  In our study, the
laser pulse is aligned with the molecular axis, has frequency
$\omega=0.057$ a.u. and a 10 cycle duration plus a 2 cycle
turn-off---we use a cos pulse with a cos$^2$ turn-off. After setting
up the initial state, we transform to a new system of coordinates,
the so called ``regularized" coordinates \cite{regularized}. This
transformation  explicitly eliminates the Coulomb singularity
\cite{Emmanouilidou2008,Emmanouilidou2009}. We regularize using the
global regularization scheme described in \cite{global}. Finally, we
propagate in time using the Classical Trajectory Monte Carlo
method \cite{CTMC}. (We note that to correctly account for EIDI and formation of H$^{*}$
we allow for both electrons to tunnel during propagation in a manner
similar to that outlined in \cite{Cohen}.)
We now select those trajectories leading to a
break up of H$_2$ with H$^{+}$, H$^{*}$ (where $*$ denotes an
electron in a $n>1$ quantum state) and a free electron as fragments.
To identify the electrons captured in a Rydberg n state 
 we first find the classical 
$n_{c}= 1/\sqrt{2|E|}$ where E is the potential plus kinetic energy
of the electron. Next, we assign to this classical number a quantum
one such that the relationship $((n-1)(n-1/2)n)^{1/3}\le
n_{c} \le (n(n+1/2)(n+1))^{1/3}$, derived in \cite{Mackellar}, is
satisfied.

\begin{figure}[h]
\centerline{\includegraphics[scale=0.35,clip=true]{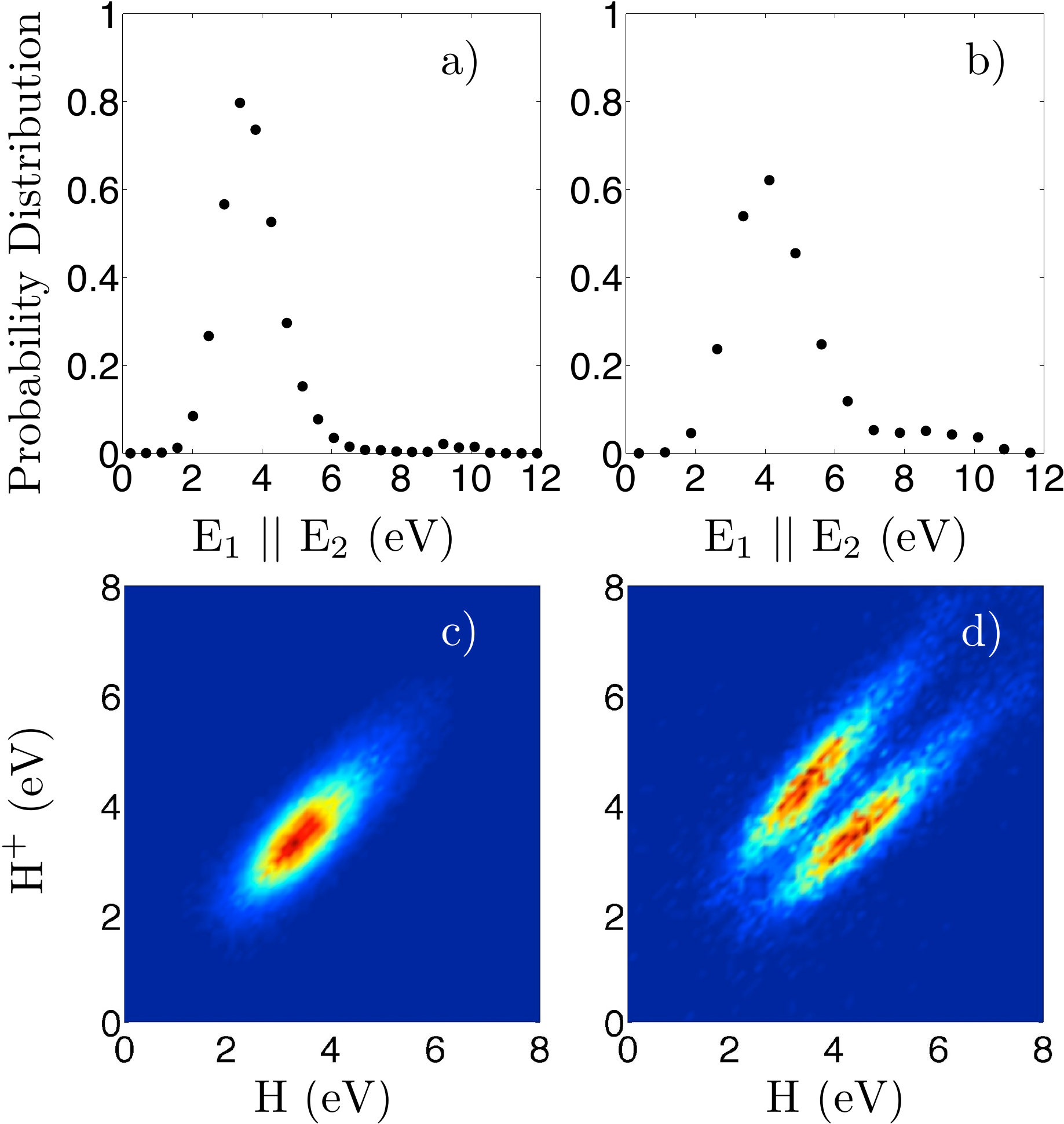}}
\caption{\label{fig:excited} (Color online) Top row: Probability distribution for either of H$^{+}$ or H$^{*}$ to have a certain amount of energy in the H$^{*}$ formation channel; Bottom row: The double differential probability in energy for H$^+$ and H$^{*}$. Left column at 1.5$\times 10^{14}$ Watts/cm$^2$ and right column at 2.5$\times 10^{14}$ Watts/cm$^2$. }
\end{figure}

As a result of our calculations we
find that this H$^{*}$ ionization channel in the strongly driven H$_{2}$ accounts for roughly 7-10\% 
of the break-up events. We compute the probability distribution as a
function of the kinetic energy of
either one of the fragments H$^{+}$, H$^{*}$ for 1.5$\times 10^{14}$
Watts/cm$^2$ and 2.5$\times 10^{14}$ Watts/cm$^2$, see
\fig{fig:excited} a) and b). We find that the maximum of the
probability distribution is around 3.5 eV. This is in very good
agreement with the experimental results in \cite{Eichmann} and
suggests that our model accurately describes this process. (The
experimental peak at about 0.5 eV due to bond-softening is not
addressed by the current work). Furthermore, we find that
the
distribution of principal quantum numbers $n$ in H$^{*}$, peaks 
around  $n=8$ resembling results for the atomic case
\cite{Eichmann2}.

% \begin{figure}[h]
%\centerline{\includegraphics[scale=0.3,clip=true]{Figure2}}
%\caption{\label{fig:excitednew} Probability distribution for the formation of highly excited neutral H with a quantum state n for a) 1.5$\times 10^{14}$ Watts/cm$^2$ and b) for 2.5$\times 10^{14}$ Watts/cm$^2$.}
%\end{figure}

 In a first attempt towards
understanding, in \cite{Eichmann}  the formation of H$^{*}$ was
attributed  to one electron effects, i.e., ``frustrated" (tunnel) ionization of the electron in
H$_{2}^{+}$---assuming the first electron ionizes immediately. ``Frustrated" ionization refers to the process where after tunneling an electron quivers in the laser field; when the field is turned off the electron does not have enough drift energy to overcome the Coulomb potential and gets captured in a Rydberg state.  Our full classical calculation shows this conjecture to be partly true. It reveals that two electron effects pervade the two pathways that we find underlying the formation of H$^{*}$. In the pathway we refer to as first,  while the first electron---the one that tunnels in the initial state---quivers in the laser field, the initially bound second electron
gains energy and ionizes. The second electron can gain energy through two mechanisms  a) a non-sequential one, where the first tunnel electron transfers energy 
through a re-collision;  b) a sequential one, where the
the initially bound
electron, independently of the first tunnel electron,  ionizes at an intermediate distance of the nuclei through
enhanced ionization of H$_{2}^{+}$.  Both for a) and b) 
  the first electron eventually undergoes ``frustrated" ionization. In both cases the
timing of the ionization of the second electron determines the
kinetic energy of the fragments.
  
  In the second pathway, the initially bound second electron gains energy through two mechanisms a) a non-sequential one where it undergoes a re-collision with the 
 quasi-free first electron;  b) a sequential one where the first electron ionizes fast and the bound electron gains energy through the lowered potential due to Coulomb explosion of the nuclei.
 Both for a) and b) the initially bound electron undergoes ``frustrated" ionization.

\begin{figure}[h]
\centerline{\includegraphics[scale=0.35,clip=true]{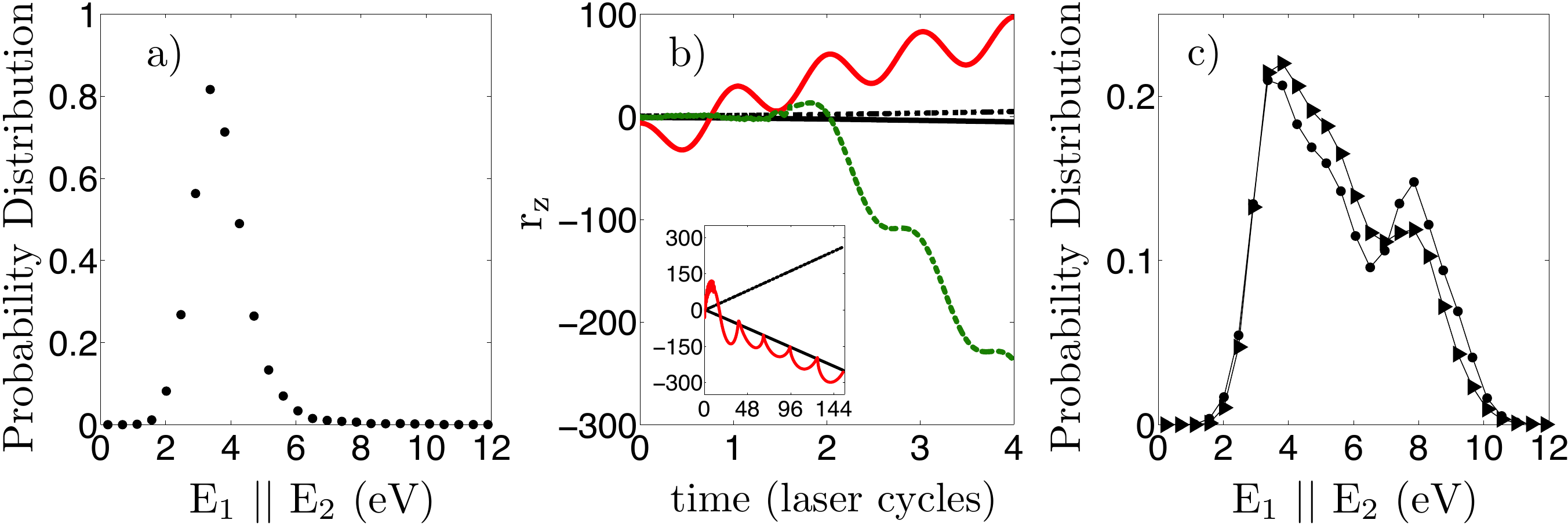}}
\caption{\label{fig:HENH-NS} (Color online) For 1.5 $\times 10^{14}$ Watts/cm$^2$ a) Probability distribution for either H$^{+}$ or H$^{*}$ to have a certain amount of energy in the non-sequential part of the first pathway of H$^{*}$; c) Probability distribution
for either of the H$^{+}$ fragments to have a certain amount of energy in the NSDI process: circles show the total probability distribution while the triangles show the contribution from the Delayed pathway; b) A typical trajectory for the non-sequential first pathway of H$^{*}$ formation with the solid/dashed  black line depicting the nucleus escaping in the same/opposite direction of initial tunneling, the dashed green line the bound electron and the red line the first electron.  In the inset of \fig{fig:HENH-NS} b) we show how the first electron undergoes ``frustrated" ionization. }
\end{figure}

\begin{figure}[h]
\centerline{\includegraphics[scale=0.35,clip=true]{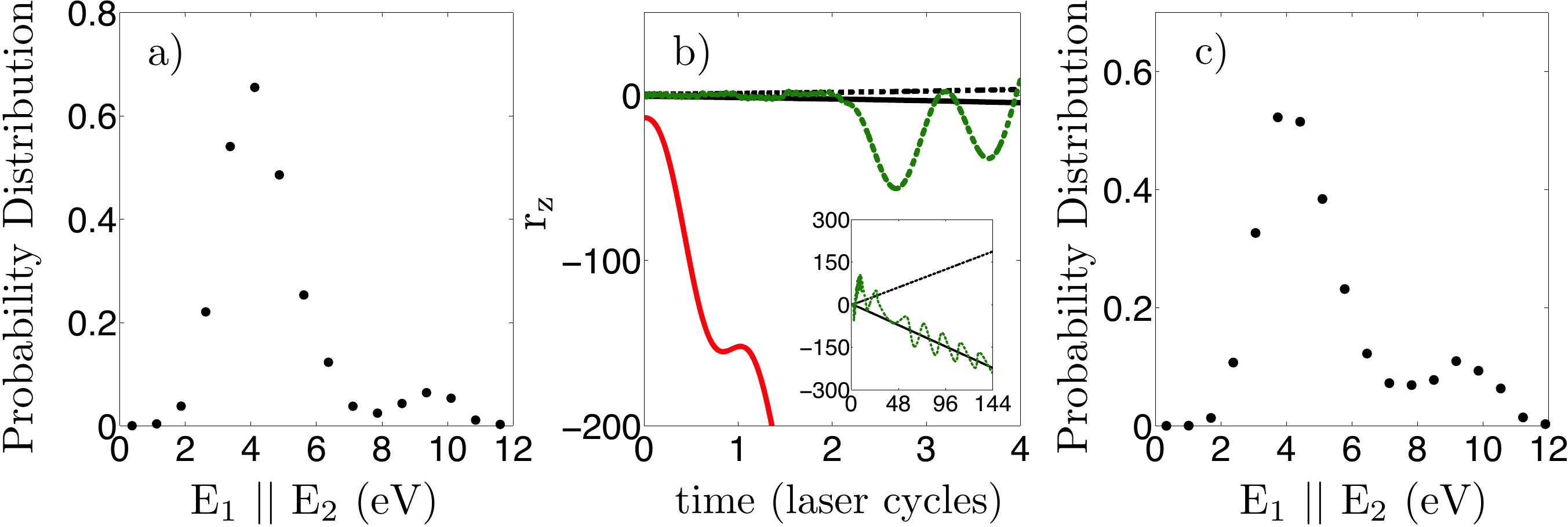}}
\caption{\label{fig:HENH-EI} (Color online) For 2.5 $\times 10^{14}$ Watts/cm$^2$ a) same as \fig{fig:HENH-NS} a) but for the sequential part of the second pathway of H$^{*}$ formation; c) Probability distribution
for either of the H$^{+}$ fragments to have a certain amount of energy in the EIDI process; b) same as \fig{fig:HENH-NS} b) but for the sequential part of the second pathway. In the inset of  \fig{fig:HENH-EI} b) we show how the second electron undergoes ``frustrated" ionization.}
\end{figure}

Two-electron effects mostly prevail in the first pathway. Indeed, in this pathway the first electron quivers in the proximity of the nuclei before the bound electron ionizes. However, in the second pathway the first electron ionizes early on while excitation of the bound electron follows only afterwards. A finding corroborating the significant role of  two-electron effects in the first pathway for small intensities is that the contribution of the first pathway decreases from 7\% of all break-up events for 1.5 $\times 10^{14}$ Watts/cm$^2$ to half that  for 2.5 $\times 10^{14}$ Watts/cm$^2$. This is consistent with the diminishing role of re-collisions 
with increasing intensity. On the other hand the second pathway's contribution remains constant around 4\% for both intensities. 
Moreover, using as a rough criterion the presence of a re-collision (minimum in the distance of the two electrons) we can further separate the trajectories of the first and second pathways to sequential and non-sequential ones. We find that the non-sequential/sequential trajectories correspond to the first electron tunneling at positive/negative phases of the laser field. While non-sequential trajectories prevail for 1.5 $\times 10^{14}$ Watts/cm$^2$ sequential ones prevail for 2.5 $\times 10^{14}$ Watts/cm$^2$. Typical trajectories of the first and second pathway are shown in \fig{fig:HENH-NS} b)
and \fig{fig:HENH-EI} b), respectively.

From the above we conjecture that the non-sequential part of the first pathway of H$^{*}$ formation  for 1.5 $\times 10^{14}$ Watts/cm$^2$ is the one that resembles the most non-sequential double
ionization (NSDI)---for H$^{*}$ double
ionization is frustrated and the first electron remains captured. Indeed,
comparing \fig{fig:HENH-NS} a)  with
\fig{fig:HENH-NS} c) we find that the probability distribution for
the energy of H$^{+}$ or H$^{*}$ in the 
non-sequential part of the first pathway of H$^{*}$ is similar to
the probability distribution for the energy of the two H$^{+}$
fragments in the Delayed pathway of NSDI. We find that NSDI is roughly 2\% of all
break-up events. As is well known, the Delayed pathway---one electron ionizes with a delay, more than a quarter laser cycle,
after re-collision---is one of the major energy transfer pathways in NSDI.
 (For details on the separation of NSDI pathways
see \cite{Emmanouilidou2009}.)  A comparison between \fig{fig:HENH-NS} a)
and \fig{fig:HENH-NS} c) further shows that in the first pathway of
H$^{*}$ formation the bound electron ionizes later compared to its  ionization time in the Delayed pathway. Indeed, larger
ionization times correspond to the nuclei Coulomb exploding at a later time,  resulting in final energies of the nuclei smaller for H$^{*}$ compared to NSDI.

The sequential part of the second pathway of H$^{*}$ formation for 2.5 $\times 10^{14}$ Watts/cm$^2$ is the one resembling the most
double ionization through enhanced ionization (EIDI)---for H$^{*}$ double
ionization is frustrated and the second electron remains captured. As
is well known (see \cite{Niikura}) in the EIDI pathway the first
electron ionizes fast while the initially bound one ionizes at an
intermediate distance of the nuclei. We find that the EIDI process
accounts for more than 50\% of all events in agreement with previous
results \cite{Niikura}. For EIDI the probability
distribution for the energy of H$^{+}$ shown in \fig{fig:HENH-EI} c)
peaks at 3.5 eV, also roughly in agreement with previous results,
see for example \cite{Madsen}. A comparison of \fig{fig:HENH-EI} a)
with \fig{fig:HENH-EI} c) shows that indeed the sequential part of the second pathway of H$^{*}$
formation resembles  EIDI.

We have now established the pathways of energy transfer underlying
H$^{*}$ formation. Next,  we seek further insight into the underlying processes by
considering the energy sharing between H$^{+}$ and H$^{*}$, see
double energy differentials in \fig{fig:excited} c) and d). We find
that H$^{+}$ and H$^{*}$ share the energy equally for 1.5$\times
10^{14}$ Watts/cm$^2$ and unequally for  2.5$\times 10^{14}$
Watts/cm$^2$, albeit the difference in energy sharing is small. The
equal energy sharing is in very good agreement with the experimental
results in \cite{Eichmann} for an intensity of 3$\times 10^{14}$
Watts/cm$^2$. However, the asymmetry we predict in energy sharing at
2.5$\times 10^{14}$ Watts/cm$^2$ suggests that the experimental
results have contributions from smaller intensities.

\begin{figure}[h]
\centerline{\includegraphics[scale=0.3,clip=true]{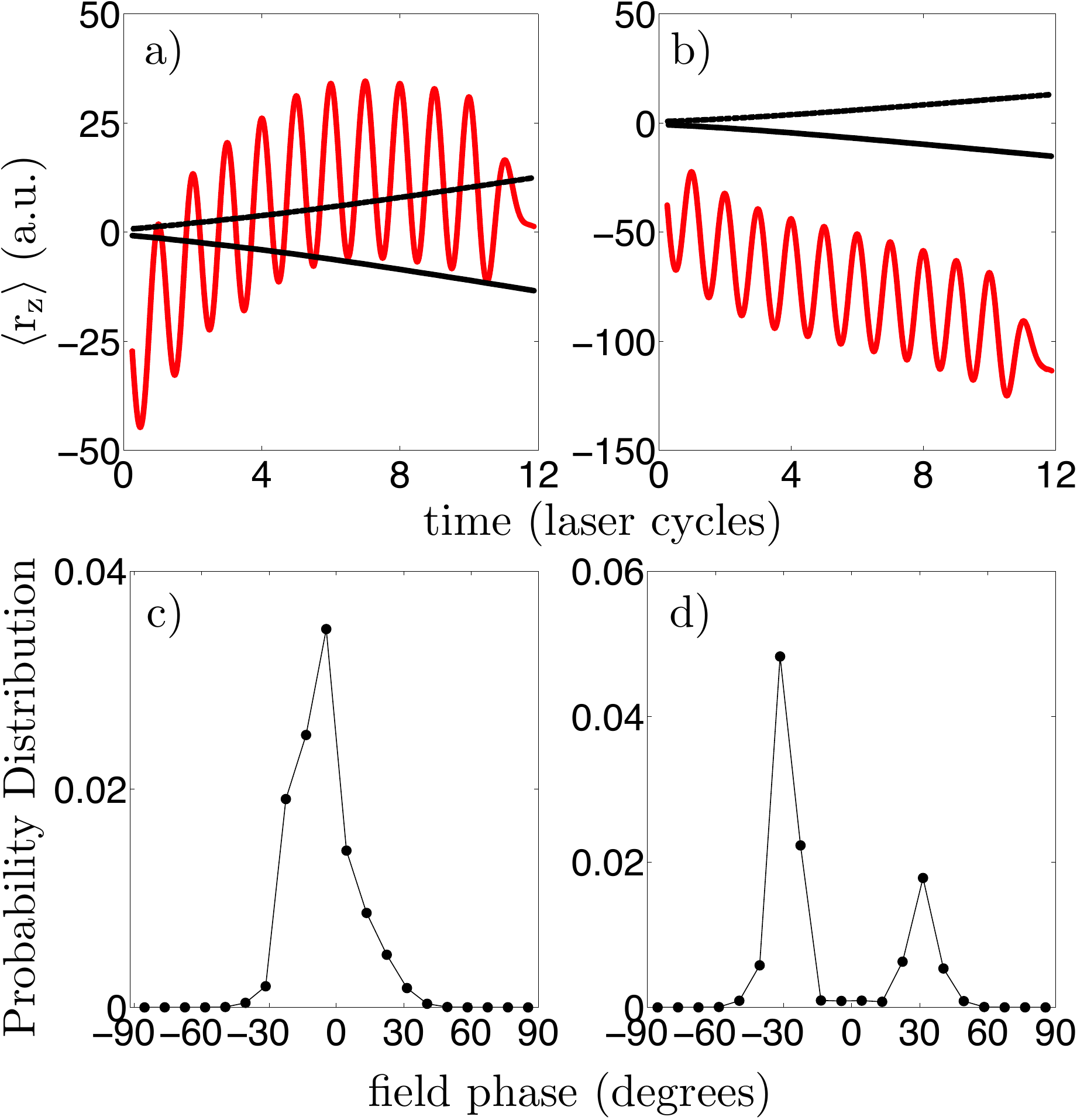}}
\caption{\label{fig:mean} (Color inline) Top row: For the first pathway, mean value of the position component z of nuclei and electrons when $z>0$ for the second electron; lines are the same as for \fig{fig:HENH-NS} b). Bottom row: Probability distribution for both pathways for the phase of the laser field when the first electron tunnels. Left/right  column for 1.5$\times 10^{14}$/2.5$\times 10^{14}$ Watts/cm$^2$.}
\end{figure}

We find the transition from symmetric to asymmetric energy sharing
to be related to the phase of the field at the time the first
electron tunnels. For 1.5 $\times 10^{14}$
Watts/cm$^2$ the distribution of the phase of the laser field when
the first electron tunnels is centered around zero, see
\fig{fig:mean} c). According to the three-step model, for phases around $0^{\circ}$ the first electron returns close to the
nuclei. Indeed, as shown in \fig{fig:mean} a), the first
electron oscillates between the two nuclei in the z (field)
direction for several periods before it
finally gets captured in a Rydberg state of H. The return of the
first electron close to the nuclei favors transfer of energy
through the non-sequential H$^{*}$ mechanism, indeed the case for
1.5 $\times 10^{14}$ Watts/cm$^2$ as discussed above.  Moreover,
with the first tunnel electron oscillating between the two nuclei, the
nuclei experience the same attractive force from the electron.  This
force combined with the repulsive force due to Coulomb explosion,
results in a symmetric energy sharing of H$^{+}$ and H$^{*}$, see
\fig{fig:excited} c).

For  2.5 $\times 10^{14}$ Watts/cm$^2$ the distribution of the phase
of the laser field when the first electron tunnels is centered
mostly around -30$^{\circ}$, see \fig{fig:mean} d). For large negative
phases the first electron does
not return close to the nuclei. Instead, it
remains on the same side that it tunneled from in the z direction before it gets captured,
 see \fig{fig:mean} b). This behavior does
not favor re-collisions and is consistent with our 
finding of sequential trajectories prevailing the formation of 
H$^{*}$
at this higher intensity. Moreover, the behavior of the
first electron causes an asymmetry in the force experienced by
the Coulomb exploding nuclei. The nucleus escaping along the same
direction as the first electron experiences from this electron
an attractive force that adds to the force due to Coulomb explosion.
However, the nucleus escaping in the opposite z direction
experiences an attractive force from the first tunnel electron that
decreases the force from Coulomb explosion. This explains the
asymmetry in energy sharing for 2.5 $\times 10^{14}$ Watts/cm$^2$ in
\fig{fig:excited} d).

 In conclusion, we have shown that
the H* formation in the break-up or Coulomb explosion of H$_{2}$ in strong
laser fields proceeds through two pathways. One where the first electron undergoes ``frustrated" ionization while the initially bound electron ionizes 
 by absorbing energy
from the field or via a collision process. In the other pathway, the first electron ionizes and the second electron gains energy either from the field or the first electron and undergoes ``frustrated" ionization.
Furthermore, we have
shown that for higher intensities electron tunneling at
large negative phases gives rise to asymmetric energy sharing
between H$^{+}$ and H$^{*}$. This asymmetry might not be possible to predict using the BO approximation since the latter assumes
that the nuclei are nearly fixed with respect to the electronic motion. This could be checked quantum mechanically with  H$_{2}^{+}$ since the asymmetry in energy sharing is mostly a one electron effect.

%is underlied by a sequential mechanism that resembles double
%ionization through enhanced ionization and a non-sequential one that
%resembles non-sequential double ionization. The former process
%prevails with increasing intensity.  Furthermore, we have shown that
%for higher intensities the influence of the electronic motion on the
%nuclear one, as a result of the electron tunneling at large negative
%phases, gives rise to asymmetric energy sharing between H$^{+}$ and
%H$^{*}$. This asymmetry is a striking demonstration of the large
%effect the motion of a light particle, such as an electron, can have
%on the motion of heavy ones, such as nuclei.

{\it Acknowledgments.} AE acknowledges support from EPSRC under grant no. EPSRC/H0031771,
 from NSF under grant no. NSF/0855403 and Teragrid computational resources under grant no. PHY110017.
We are also grateful for discussions with P. Corkum, A. Staudte, A. Saenz and S. Patchkovskii.

\end{document}